\documentstyle[preprint,aps,epsfig]{revtex}
\newcommand{\ignore}[1]{}

\begin{document}
\draft
\title{Experimental investigation of universal parametric
correlators using a vibrating plate}

\author{K. Schaadt$^{1,2}$ and A. Kudrolli$^1$}

\address{$^1$ Department of Physics, Clark University,
              Worcester, MA 01610, U.S.A.\\
         $^2$ Center for Chaos and Turbulence Studies,
           Niels Bohr Institute, Blegdamsvej 17,
              2100 Copenhagen \O, Denmark.\\}

\date{\today} 

\maketitle

\begin{abstract}

The parametric variation of the eigenfrequencies of a chaotic plate
is measured and compared to random matrix theory using recently
calculated universal correlation functions. The sensitivity of the
flexural modes of the plate to pressure is used to isolate this
symmetry class of modes and simplify the data analysis. The size of
the plate is used as the external parameter and the eigenvalues are
observed to undergo one or two oscillations in the experimental
window. The correlations of the eigenvalues are in good
agreement with statistical measures such as the parametric number
variance, the velocity autocorrelation, and the intralevel velocity
autocorrelation derived for the Gaussian Orthogonal Ensemble of
random matrix theory. Our results show that the theory can be also 
applied to wave systems other than quantum systems.

\end{abstract}



\narrowtext

It has been widely recognized that the eigenvalues of a quantum
system show universal features that depend only on the presence or
absence of chaos in the corresponding classical or ray
system~\cite{Casati:1995,Mehta:1990}. For example, it has been
established that the eigenvalues of integrable systems display
Poisson statistics, and chaotic systems with time-reversal symmetry
show statistics which are similar to the Gaussian Orthogonal
Ensemble (GOE) of Random Matrix Theory (RMT)~\cite{Bohigas:1984}.
The universality has been confirmed using not only quantum systems,
but also systems which obey an elastomechanical wave 
equation~\cite{Weaver:1989,Ellegaard:1995,Ellegaard:1996}. The
difference in the statistical properties has been recognized 
to be due to the presence of level
repulsion which were demonstrated as avoided crossings as a system
parameter was varied. However, it was postulated only recently that
the resulting fluctuation of the energy levels also show universal
properties which are independent of the nature of the
parameter~\cite{Simons:1993a,Simons:1993b}.

When a quantum system is subjected to a perturbation via an external
parameter $X$, the eigenvalues change and oscillate as a
function of $X$. Using supersymmetry techniques, Simons and
Altshuler~\cite{Simons:1993a,Simons:1993b} were able to calculate the
correlations as a function of external parameter for energy levels with
Wigner-Dyson distributions of RMT. The agreement of their analytical
results with numerical simulations of disordered metallic rings and a
chaotic billiard led them to the remarkable conjecture that
correlations in the eigenvalues show universal features which
are {\em independent} of the nature of the perturbation after
appropriate normalization. Here the proper rescaling required to
compare across different systems is given by expressing the energy
$E$ in units of the local mean level spacing $\Delta$, and the
parameter in units of the square root of the local mean squared
slope:
\begin{equation}
\varepsilon = E/\Delta \;\;\;\;\;\; 
x = \sqrt{\left< \left({d\varepsilon \over dX}\right)^2 \right>} X \,
\, ,
\label{x_def}
\end{equation}
where $\varepsilon$ is the normalized energy, and $x$ is the
rescaled external parameter.

The conjecture was tested further with numerical simulations of a
Hydrogen atom in a magnetic field, where agreement was found over a
certain parameter range, but systematic deviations were also found
because the system is only partially chaotic~\cite{Simons:1993c}.
Although some of the correlations have been indirectly 
tested in the conductance fluctuations of electrons in ballistic 
cavities~\cite{Sivan:1994,Huibers:1998}, and also in microwave 
cavities~\cite{Barth:1998} and quartz blocks~\cite{Ellegaard:1998} where
bouncing ball-like modes complicates the analysis,
there has been no report of a direct experimental test of their universality.

In this paper, we report direct experimental evidence for the
universality of the above mentioned parametric correlators. A
freely vibrating plate with the shape of a
Sinai-Stadium~\cite{Kudrolli:1994} is used and the smooth motion of
the eigenfrequencies is measured as a function of the size of the
plate. Two classes of uncoupled modes exist in an isotropic plate:
{\it flexural}, for which the displacement is perpendicular to the
plane of the plate, and {\it in-plane}, for which the
displacement is in the plane of the
plate~\cite{Graff:1975,Hugues:1997}. We are able to experimentally
isolate the flexural modes and therefore can simplify the analysis
by not having to consider problems associated with mixed
symmetries. The flexural modes obey a scalar equation for the
displacement $W$ perpendicular to the plate:
\begin{equation}
(\nabla^2 - k^2)(\nabla^2 + k^2)W = 0 \;\;\;,
\label{flex_eq}
\end{equation}
where $k$ denotes the wavenumber. The dispersion relation is given by
\begin{equation}
f = {k^2 \over 2\pi} \sqrt{E_Y h^3 \over 12\rho(1-\nu^2)}\;\;\;,
\label{flex_disp_rel}
\end{equation}
where $f$ is the frequency, $h$ is the thickness of the plate,
$\rho$ is the density, $E_Y$ is Young's modulus, and $\nu$ is
Poisson's ratio.
Any solution $W$ of Eq.\ (\ref{flex_eq}) can be written as a superposition of
two modes, $W_1$ and $W_2$, where
\begin{equation}
(\nabla^2 + k^2)W_1 = 0 \;\;\;\;\, \mbox{and} \;\;\;\; 
(\nabla^2 - k^2)W_2 = 0 \;\;.
\end{equation}
$W_1$ is a solution to the Helmholtz equation with free
boundary conditions. $W_2$ is an {\it exponential} mode or {\it boundary}
mode. The boundary modes are responsible for only about one percent
of the density of states~\cite{Bogomolny:1998,Hugues:1997} and do not appear 
to alter  the universality of the eigenvalues. 
Equation (\ref{flex_eq}) is an approximation to the full elastomechanical 
wave equation in the limit where the wavelength is much larger than the
thickness of the plate. The typical wavelength in our experiments is 8 mm,
and the thickness of the plate is 2 mm. In this case, Eq.\ (\ref{flex_eq}) is
a good approximation.

The eigenvalue statistics is first confirmed to agree very well with 
GOE statistics using traditional measures such as the spacing 
statistics $P(s)$ and the spectral rigidity $\Delta_{3}(L)$, 
demonstrating the quality of the data and the universality of the 
geometry. We then compare statistical observables of the eigenvalue
motion as a function of the parameter to analytical calculations.
In particular, we find that the data agrees with calculations of
the parametric number variance 
$v(x)$ by Simons et al.~\cite{Simons:1993c} and shows 
a linear behavior for small $x$ which is different from semiclassical
calculations~\cite{Goldberg:1991}. To investigate correlations in
energy-parameter space, comparisons are made with the exact
calculations for the intralevel velocity autocorrelation $\tilde
c(w, x)$ which describes the correlations between the rate of
change of eigenvalues separated in energy by
$w$ and in parameter by $x$~\cite{Simons:1993b}. Good agreement is
observed for selected values of $w$ and over all $x$. Another
statistical measure is the velocity autocorrelation $c(x)$ which
correlates the rate of change of eigenvalue as a function of
parametric separation $x$. For this quantity we find that the data
is in good agreement at small and intermediate values of the
parameter. Deviations are observed for larger
$x$ where statistical sampling is poor. 
Combined, these results provide the first experimental evidence for 
the universality of a broad class of the statistical observables of
parametric level motion that have been studied theoretically.  

\begin{figure}
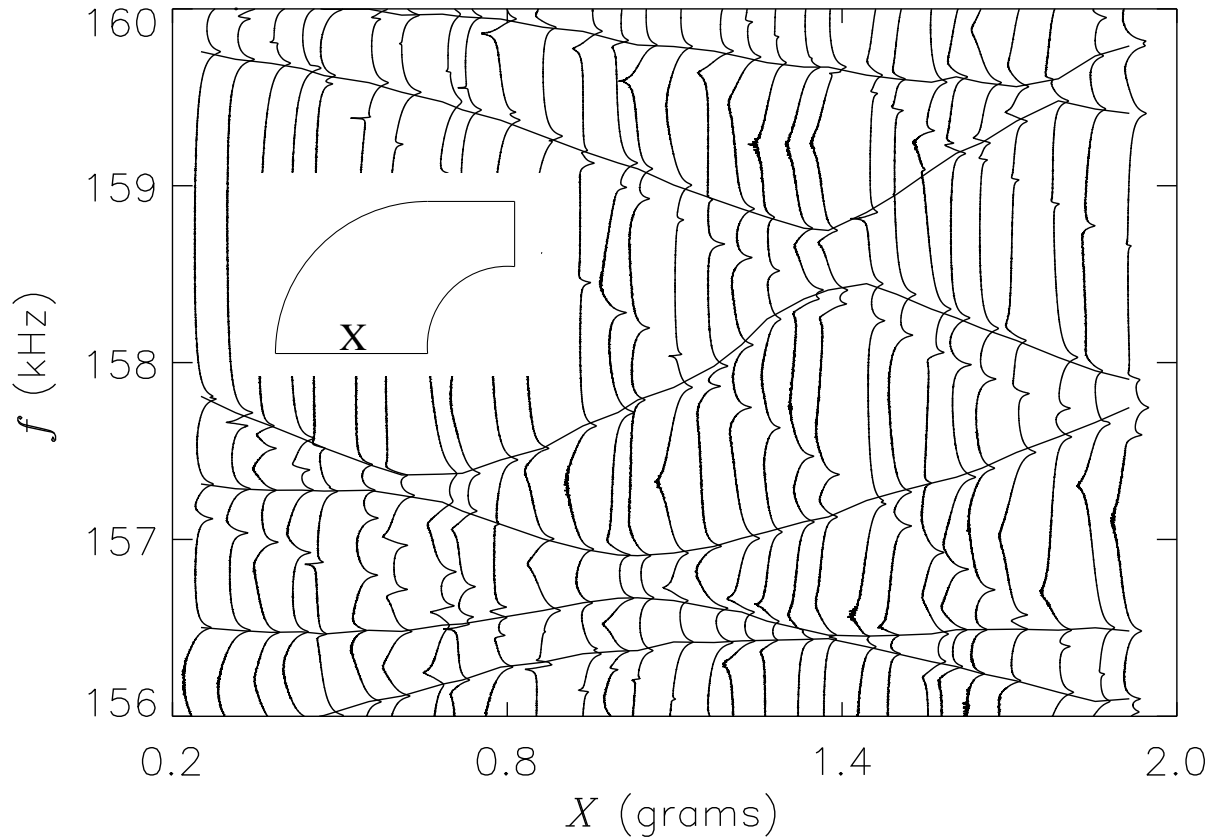

\caption{\footnotesize The transmission amplitude as a function of
the frequency in steps of the parameter $X$. The flexural modes are joined by a solid curve to 
guide the eye. Other modes pass through the diagram without any interaction with the flexural
modes. These are the in-plane modes, which are not included in the data analysis (see text). 
Inset: The shape of the Sinai-Stadium plate. The side which is polished to effect a
parametric change is indicated by $X$.}
\label{plate}
\end{figure}

In the experiments we use an Aluminium plate of thickness 
2.0 mm, machined in the shape of a quarter Sinai-Stadium with radii
40 mm and 70 mm (see Fig.~\ref{plate}). The plate rests on three
piezoelectric transducers,  of which one is a transmitter and two
are receivers.  We measure acoustic transmission spectra of the
plate using a HP 4395A network analyzer. A sample of the
transmission signal at different values of the parameter is shown
in Fig.~\ref{plate}. The amplitude of the resonances depend on  the
location of the transducers but the eigenfrequencies are unchanged.
The plate is kept in a temperature controlled oven held at 
300\, K to within 1 mK. A vacuum system ensures that the air
pressure is below $10^{-1}$\, Torr, which is low enough that air
damping of the plate is insignificant compared to other damping
mechanisms. Of the two classes of modes, flexural modes are more
sensitive to the presence of air damping than in-plane modes 
because of the flexural out-of-plane oscillation. We find that
going from vacuum to atmospheric pressure, the $Q$-factor of the
flexural modes decreases by at least a factor of 3, whereas the
$Q$-factor for the in-plane modes is unchanged. 

We first measure the transmission spectrum of the plate, then
decrease the size of the plate by sanding off material at the
longest straight edge, as indicated in Fig.~\ref{plate}. The amount
of material removed is determined by measuring the mass of the
plate to within
$5\times 10^{-5}$\, grams. Approximately $5\times 10^{-2}$\, grams
is removed each time and in all 6\% of the material is removed in 63
steps.  The spectrum is measured in the interval between 100\, kHz and 
300 kHz.  Periodically, the spectrum is also measured at 1 atmosphere 
to identify the flexural modes.  After this separation, we find 
approximately 300 resonances, of which 25 drift out of the frequency 
window due to the overall increase in frequency when the size of the 
plate is decreased.  A resonance frequency can be determined to within 
0.5 Hz by fitting the resonance peak to a Breit-Wigner function.  We 
are confident that all eigenfrequencies in the frequency window are 
detected, because it is impossible for the amplitude of a resonance 
peak to lie below our detection level for all 63 values of the 
parameter.  The absence of interaction of the flexural modes with the 
in-plane modes is checked to within experimental accuracy by noting 
a lack of interaction at flexural-in-plane encounters (see 
Fig.~\ref{plate}).

In the data analysis, the implementation of the normalization or
{\it unfolding} given by Eq.~(\ref{x_def}) is of great importance. 
Since the cumulative level density or {\it staircase function} for 
a freely vibrating plate was recently 
calculated~\cite{Bogomolny:1998}, both the mean level spacing and
the mean squared velocity are known analytically. This knowledge
can be directly applied to our data, which makes the data analysis 
very clean from a theoretical viewpoint.

\begin{figure*}
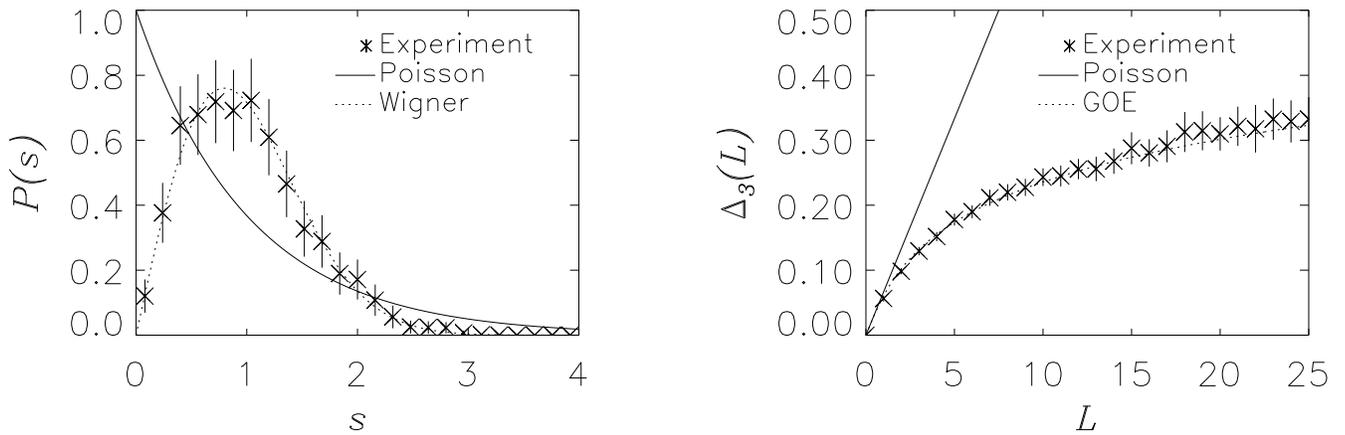

\caption{\footnotesize Left: Distribution of nearest neighbor
spacings $P(s)$ from all 63 measured spectra (crosses), compared to the
Wigner distribution of RMT (solid curve). Right: Spectral rigidity $\Delta_3(L)$ for the 
experimental data (crosses) compared to the GOE result (solid curve).}
\label{spec_stat}
\end{figure*}

We start by presenting our result for the distribution of nearest
neighbor spacings $P(s)$ and the spectral rigidity $\Delta_{3}(L)$
which are shown in  Fig.~\ref{spec_stat}. We find complete
agreement with RMT for both observables.  Fully chaotic systems are
very rare and most chaotic geometries  have regions in phase space
which are integrable. The Sinai-Stadium geometry is no exception
and is known to have small regions of integrability. However if
these regions are very small, they can support an integrable level
only at very high frequencies, and therefore complete agreement
with GOE is expected and observed.

\begin{figure}
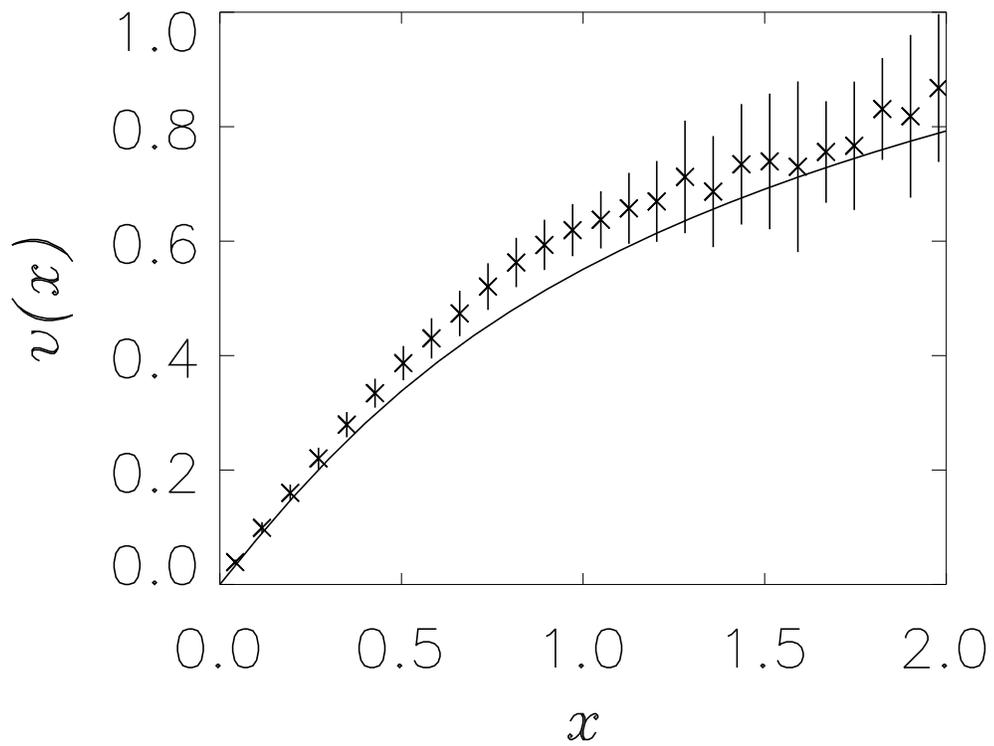

\caption{\footnotesize Comparison of experimental data (crosses)
and RMT (solid curve) for the parametric number variance $v(x)$.}
\label{par_num_var}
\end{figure}

We now present the main results which is the
correlations in the parametric variation of the eigenfrequencies.
The parametric number variance $v(x)$ is defined as:
\begin{equation}
v(x) = \langle (n(\varepsilon,x')-n(\varepsilon,x'+x))^2 \rangle \;\;,
\label{par_num_var_def}
\end{equation}
where the average is over the parameter $x'$ and energy
$\varepsilon$. Here, $n(\varepsilon,x)$ is the staircase
function which  counts the number of energy levels at fixed $x$
with energy lower than $\varepsilon$. The parameter $x$ has been
normalized according to Eq.~(\ref{x_def}), as explained above. The 
variance measures
the difference in the number of eigenvalues which are below a fixed
value of normalized energy $\varepsilon$. Therefore this quantity
measures the collective motion of levels under parametric
change~\cite{Goldberg:1991}. The comparison of the data with the
theory is shown in Fig.~\ref{par_num_var}. The $v(x)$ calculated
from the data grows linearly from zero and has a slope of $0.8 \pm
0.01$ which is in excellent agreement with the calculated value of
$\sqrt{2/\pi} \sim 0.797$ by Simons et al.~\cite{Simons:1993c}. A
saturation is expected at large values of $x$ and therefore the
$v(x)$ becomes sub-linear at higher $x$. 

However, $v(x)$ does not give an indication of the correlations
in the oscillations of the eigenvalues with the parameter $x$.
To investigate such correlations, a new set of measures are required 
that study the rate of change of eigenvalue as a function of 
parameter~\cite{Simons:1993a}. One example
is the {\it intralevel velocity autocorrelation} $\tilde
c(\omega,x)$, which correlates velocities which are separated
by a distance $x$ in parameter space and by a distance
$\omega$ in energy:
\begin{equation}
\tilde c(\omega,x) = {{\sum_{n,m}\langle
\delta(\varepsilon_n(x')-\varepsilon_m(x'+x) - \omega)
{\partial \varepsilon_n(x') \over \partial x'}
{\partial \varepsilon_m(x'+x) \over \partial x'}
\rangle}
\over {\sum_{n,m} 
\langle \delta(\varepsilon_n(x')-\varepsilon_m(x'+x) - \omega)\rangle}}
\end{equation}
The average is over the parameter $x'$. Using the
supersymmetric nonlinear $\sigma$ model developed by
Efetov~\cite{Efetov:1983}, Simons and Altshuler derived an integral
representation for the intralevel velocity autocorrelation. Another
correlation is the {\it velocity autocorrelation} $c(x)$ which
correlates velocities which belong to the same energy level:
\begin{equation} 
c(x) = \left < {\partial \varepsilon(x') \over \partial x'}
{\partial \varepsilon(x'+x) \over \partial x'}
\right >
\end{equation} 
The brackets denote an average over the parameter $x'$ and the
energy
$\varepsilon$. For this correlator no analytical results exist for
intermediate values of $x$. Therefore we compare our result for
$c(x)$ to a curve calculated by Mucciolo~\cite{Mucciolo:1996} using
large GOE matrices which agrees with the analytical results in the
limit of large and small $x$. 

\begin{figure*}
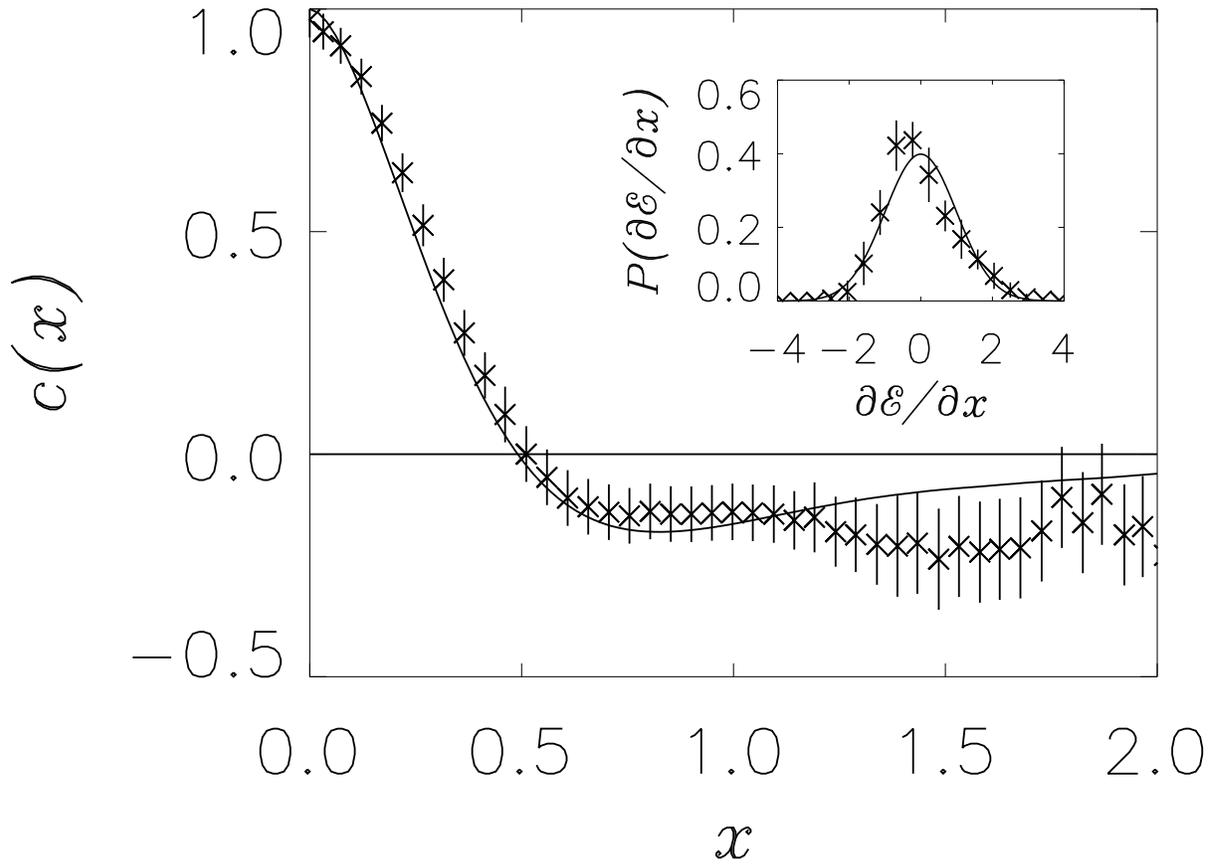

\caption{\footnotesize The velocity autocorrelation $c(x)$ (crosses)
compared to RMT calculations (solid curve). Inset: Distribution of eigenvalue
velocities compared to a Gaussian distribution.}
\label{vel_corr}
\end{figure*}

We first present the result for the velocity autocorrelation
$c(x)$ (see Fig.~\ref{vel_corr}). For values of $x$ smaller than 1,
we find good agreement with the numerical RMT
curve~\cite{Mucciolo:1996}. At larger values of $x$, however, we
see a deviation which is outside the experimental error bars.
The shape of the correlation function indicates
that the slope
${\partial \varepsilon(x)/\partial x}$ changes smoothly and
has opposite signs near $x = 0.5$ because the parameter $x$ has been
normalized to correspond to approximately one oscillation for
$x=1$. This behavior of the
correlation functions indicates that, locally, there is a particular 
length scale over which eigenfrequencies oscillate. The distribution 
of velocities ${\partial
\varepsilon(x)/\partial x}$ of the eigenvalues should be a Gaussian
with a mean value of zero. The data is shown in the inset to
Fig.~\ref{vel_corr}. The data is close to a Gaussian, but is
slightly asymmetric with more velocities of small magnitude which
are negative than positive. We emphasize that the mean slope is
zero, indicating that this discrepancy does
not originate in the normalization of the eigenfrequencies. 
We believe that the deviation is due to a
finite data set. It appears that the correlations are very robust
and give good agreement even if the velocity distribution is not
exactly Gaussian.

\begin{figure}
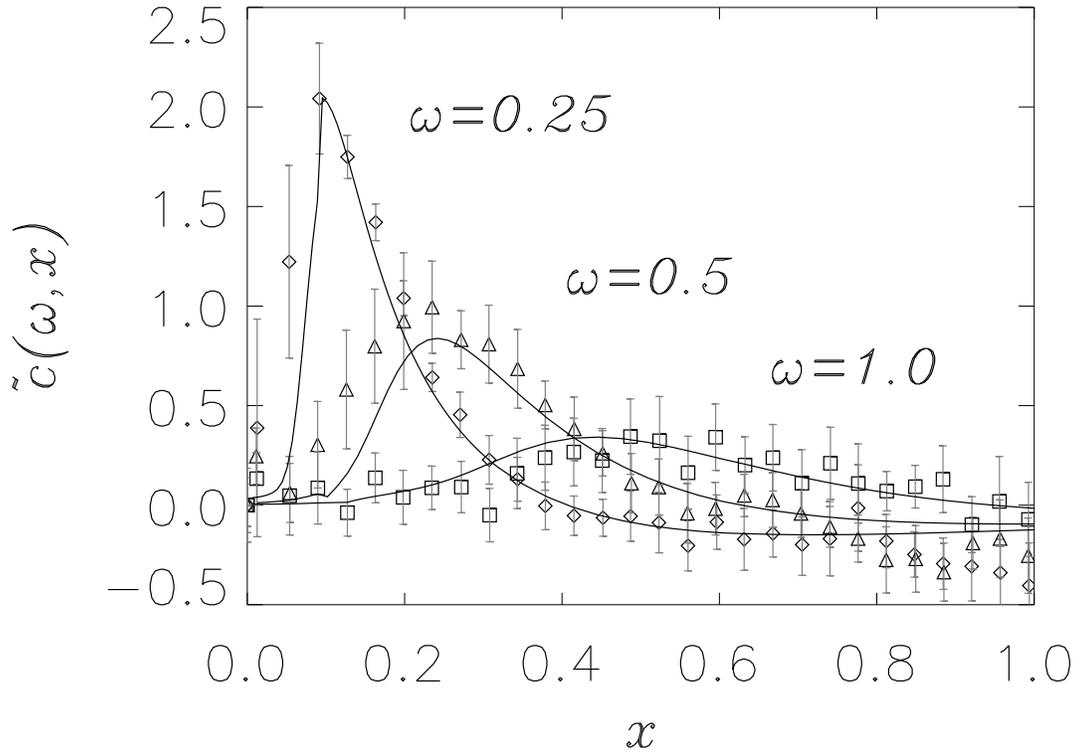

\caption{\footnotesize The intralevel velocity autocorrelation for
$\omega$=0.25 (diamonds), $\omega$=0.5 (triangles), and $\omega$=1.0 (squares). 
The theory curves correspond to
analytical calculation of Simons and Altshuler [8] using
supersymmetry techniques.}
\label{intralevel_vel_corr}
\end{figure}

To make a more stringent test of the correlations, we compare our
data with the intralevel velocity autocorrelation $\tilde
c(\omega,x)$ for $\omega = 0.25$, $\omega =
0.50$, and $\omega = 1.0$ as shown in
Fig.~\ref{intralevel_vel_corr}. We compare our data to a numerical
evaluation of the integral representation of this
correlator~\cite{Simons:1993b}. In calculating these quantities we
have averaged over a small energy window of 
$\delta \omega = 0.03$ which is also done in the theoretical
calculations. The occurrence of the peaks in the correlation
functions and the systematic increase of the value of $x$ where
the peak occurs can be understood from the fact that near an
avoided crossing, one has to go across by nearly as much along the
normalized energy axis as along the parameter axis to encounter a
similar slope (see Fig.~\ref{plate}). The comparison of the data
in Fig.~\ref{intralevel_vel_corr} shows very good agreement for all
three values of $\omega$, validating the theory. 

In conclusion, we have investigated experimentally the parametric
level motion of the flexural modes of a freely vibrating plate as a
function of the size of the plate. We have used our data to
calculate statistical quantities which probe the parametric motion
of the levels, and found agreement with the universal predictions of
RMT. The agreement with RMT
suggests that the universal predictions for parametric level motion 
extends beyond quantum chaotic systems to a wider range of wave 
systems, including acoustical waves.

We thank B.\ Simons and E.\ Mucciolo for providing the theoretical data, and
M.\ Oxborrow and J.\ Norton for technical assistance. We thank O.\ Brodier and
H.\ Gould for useful discussions. This work 
was supported by the Danish National Research Council (K.S.), 
Research Corporation and by an Alfred P. Sloan Fellowship (A.K.). 
We thank Hewlett-Packard for a partial equipment grant. 



\end{document}